# Real-time Monitoring of Cellular Cultures with Electrolyte-gated Carbon Nanotube Transistors

*Francesca Scuratti,[‡#] Giorgio E. Bonacchini,[‡#†] Caterina Bossio,[#] Jorge M. Salazar-Rios,[§] Wytse Talsma,[§] Maria A. Loi,[§] Maria R. Antognazza,[#] Mario Caironi[#]\**

[#] Center for Nano Science and Technology @PoliMi, Istituto Italiano di Tecnologia, Via Giovanni Pascoli, 70/3, 20133 Milano, Italy.

[l] Department of Electronics, Information and Bioengineering, Politecnico di Milano, Piazza Leonardo da Vinci, 32, 20133 Milano, Italy.

[§] Zernike Institute for Advanced Materials, University of Groningen, Nijenborgh 4 9747 AG Groningen, The Netherlands.

ABSTRACT: Cell-based biosensors constitute a fundamental tool in biotechnology, and their relevance has greatly increased in recent years as a result of a surging demand for reduced animal testing and for high-throughput and cost-effective in vitro screening platforms dedicated to environmental and biomedical diagnostics, drug development and toxicology. In this context, electrochemical/electronic cell-based biosensors represent a promising class of devices that enable long-term and real-time monitoring of cell physiology in a non-invasive and label-free fashion, with a remarkable potential for process automation and parallelization. Common limitations of this class of devices at large include the need for substrate surface modification strategies to ensure cell adhesion and immobilization, limited compatibility with complementary optical cell-probing techniques, and need for frequency-dependent measurements, which rely on elaborated equivalent electrical circuit models for data analysis and interpretation. We hereby demonstrate the monitoring of cell adhesion and detachment through the time-dependent variations in the quasi-static characteristic current curves of a highly stable electrolyte-gated transistor, based on an optically transparent network of printable polymer-wrapped semiconducting carbon-nanotubes.

MAIN TEXT

**Introduction**

A recent trend in toxicology and drug development focuses on enhancing the effectiveness and the comprehension of drug testing experiments performed on cell cultures, with the aim of limiting the exploitation of animal models, as well as to allow for more cost-effective and up-scalable protocols.[1–3] Cell-based assays allow in fact to assess the effects of certain biochemical perturbations, and thus their potential toxicity, by monitoring cell viability upon exposure to given molecules and/or pollutants. In particular, electric cell-substrate impedance sensing (ECIS) has been effectively used to assess bioviability, demonstrating a potential for high-throughput monitoring of cells adhesion and proliferation *in vitro*.[4–7] However, these systems are generally not compatible with the microscopy techniques extensively used for life-sciences research, the current gold standard for cell physiology monitoring, because of either the opaque nature of the substrates and/or of the electrodes used for sensing. The excellent works by Róisín Owens and collaborators have in part overcome the limitations of standard ECIS, demonstrating organic electrochemical transistors able to detect and measure cell adhesion, proliferation and detachment *in vitro* with enhanced sensitivity and temporal resolution compared to standard technologies, with the added advantage of allowing for simultaneous electrical and optical analyses.[8,9] Nonetheless, the electric approaches so far reported for cell viability monitoring, whether performed with electrodes or with transistors, are still based on step-function response analysis or AC measurements, either frequency-dependent or single-frequency, with the only exception of tubular transistors for 3D cell cultures monitoring, recently proposed by Pitsalidis *et al*.[8–12] The rather complex instrumentation required for this type of measurements, along with the need for reliable and well-established equivalent electrical models of the cell-substrate impedance for data interpretation, do not favor the large-scale automation and parallelization of these techniques, along with their portability. In this work we demonstrate that a low-voltage electrolyte-gated field-effect transistors (EGFETs), based on a solution-processed network of single-walled carbon-nanotubes (SWCNT), is able to provide information on the adhesion, proliferation and detachment processes in three different cell models, through the modification of the drain-source quasi-static current of the device, without need for electrical impedance measurements and related modeling. Our approach thus favors the portability of devices for environmental monitoring and medical diagnostics applications in the lab-on-chip format, further facilitated by the ease of processability and the operating stability of the semiconductor used in this work, without compromising the compatibility with gold-standard optical probes.

The choice of polymer-wrapped single-walled SWCNTs networks as semiconducting layer in our EGFETs is motivated by a number of factors that include SWCNTs excellent electronic properties, which favor high current density levels even at low biasing voltages, thus limiting the stress on cell cultures.[13,14] The former is a critical aspect for electrical cell-based biosensors, as bias voltages within a narrow window (0.3 – 0.6 V) are required in order to avoid cell stress, damage and death.[15] Moreover, our group has previously demonstrated that it is possible to deposit inks based on such polymer-sorted SWCNTs by inkjet-printing in order to realize complementary circuits based on SWCNTs random networks. The possibility to thus pattern the semiconductor on various substrates may favor device integration in microfluidic platforms for lab-on-chip applications.[16]

The structure of our EGFETs (Figure 1a) consists in a spin-cast network of solution-dispersed polymer-wrapped SWCNT bridging drain and source electrodes. The channel of the transistor is gated by applying a potential to a third electrode immersed in an aqueous gating medium, namely the commonly used cell growth Dulbecco's modified Eagle's medium (DMEM). The difference in potential between the gate and source/drain electrodes introduces a field under which ions dispersed in the gating medium drift until they reach the semiconductor/electrolyte interface, thus establishing an electric double layer (EDL) and determining a controlled variation of the electronic charge density within the semiconductor.[17] According to the gradual channel approximation for FETs, in the linear regime the current flowing from the drain to the source electrodes through the active material ($I_{ds}$) depends on the gate potential ($V_{gate}$) as follows: $I_{ds} \propto \mu C (V_{gate} - V_{th}) V_{ds}$, where $\mu$ is the charge carrier mobility in the semiconductor, $V_{th}$ the threshold voltage of the device, $V_{ds}$ the drain-source voltage, and $C$ is the specific capacitance at the semiconductor/electrolyte interface.[17,18]

In order to achieve optimal operation of the proposed EGFETs, the choice of high electronic quality SWCNTs has to be accompanied by a suitable substrate engineering to circumvent typical issues encountered in solution-processed SWCNTs, such as pronounced hysteresis and large onset voltages, which complicate low voltage operation. To this end, we introduced hexamethyldisilazane (HMDS) surface passivation prior to SWCNT deposition to drastically reduce hysteresis and shift the threshold voltage to lower absolute values. Figure S2 reports the transfer characteristic curves ($I_{ds}$ versus $V_{gate}$) of both HMDS-treated and untreated EGFETs operating in aqueous environment: the former demonstrate a significantly lower threshold voltage ($\Delta V_{th}$ = 0.35 V) and hysteresis (the area enclosed in the loop is reduced by ≈ 65%), as well as improved cycling stability and curve ideality. As a result, the HDMS-treated devices achieve transconductances as high as 55 µS/cm at "cell-friendly" voltages (Figure 1c). Importantly, our polymer-wrapped SWCNTs EGFETs satisfy two further critical points for reliable in vitro cell monitoring. First, they are characterized by a remarkable operating stability in aqueous environments, as data in Figure S3 of the Supporting Information display. Second, EGFETs are based on optically transparent, semiconducting layers, as SWCNTs can be deposited to form few nanometers thick networks with optimal transport properties and very low optical density (Figure 1d-1e and Supporting Information).

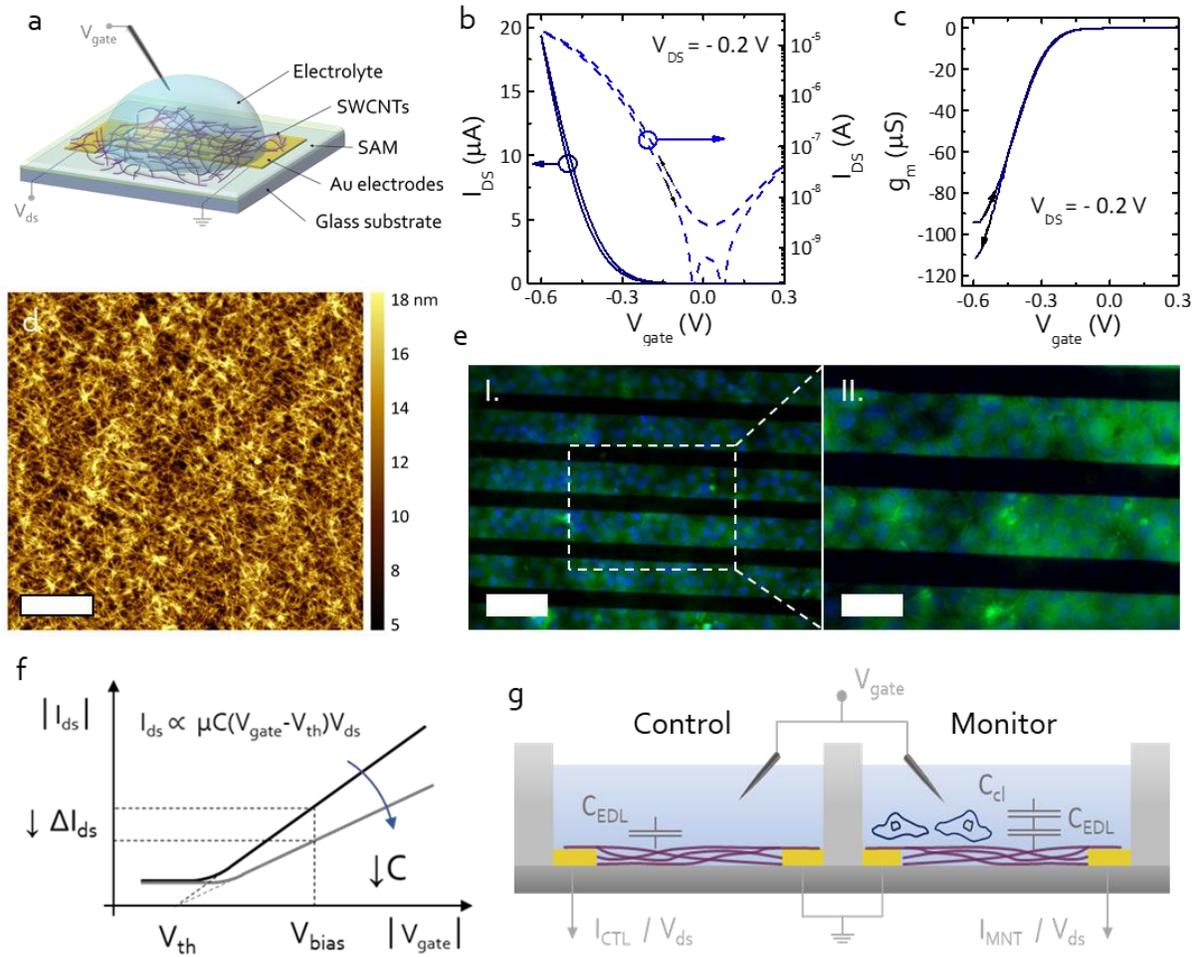

**Figure 1.** Device structure and operation. (**a**) Schematic diagram of the proposed EGFET device. The semiconducting layer consists of high pressure carbon monoxide conversion (HiPCO) SWCNTs functionalized with poly(3-dodecylthiophene-2,5-diyl) (P3DDT) as wrapping polymer. Channel length $L$= 40 μm, channel width $W$ = 20000 μm. (**b,c**) Transfer curves in semi-log (dashed) and linear (solid) scale of SWCNT based EGFETs operated in Dulbecco's modified Eagle's medium (DMEM) (**b**): the devices are biased in a gate voltage range (0.3 V to -0.6 V) compatible with cell survival, and exhibit optimal operation with limited hysteresis, low turn-on voltages and maximum transconductances (**c**) of 112 μS (calculated as $g_m = dI_{ds}/dV_g$). (**d**) AFM image of the spin-cast SWCNT layer, showing a dense interconnected network with an optimal coverage of approximately 93% (scale bar is 1 μm). (**e**) Representative immunofluorescence
6

images acquired by an inverted microscope of MDCK-II cell models cultured on EGFETs, demonstrating the device compatibility with optical probing techniques (scale bars in **I.** and **II.** are 80 μm and 40 μm respectively). (**f**) Graphical description of the impact of the variation of capacitive coupling at the semiconductor/electrolyte interface on the transistor characteristic curves. (**g**) Representation of cell viability measurement set-up, comprising a control (CTL) and a monitoring (MNT) device.

In our device, cell presence at the semiconductor/electrolyte interface can be detected through the variations in electrostatic coupling between the electrolyte and the semiconductor, as the overall capacitive contribution of the cell layer ($C_{cl}$) is added in series to the EDL capacitor ($C_{EDL}$), leading to an effective interface capacitance of:

$$C = \left(\frac{1}{C_{EDL}} + \frac{1}{C_{cl}}\right)^{-1} \tag{1}$$

Figure 1f gives a graphical representation of the operating principle of the viability measurements, where, as schematically described in Figure 1g, two equally biased EGFETs acts respectively as control (CTL) and monitoring (MNT) devices. As cellular adhesion and proliferation on the MNT device is reflected in the overall decrease in $C$, the sensing mechanism is such that the process can be monitored by simply evaluating the evolution of $I_{ds}$ versus time at a given biasing voltage, with respect to one, or eventually more, control devices to correct for eventual drifts.[4,15] This descriptive operating model is realistic only if $C_{cl}$ is relatively independent of $V_{gate}$, and if the cell cleft resistance is not significantly lower than the cell layer resistance.[19] In this work $V_{gate}$ has been limited to a voltage range that guarantees linear behaviour of the devices in the presence of cell layers adhering at the semiconductor/electrolyte interface (see Experimental Section).

**Results and discussion**

The SWCNT EGFETs have been tested as cell-proliferation monitors with three different models of adherent cells, namely Homo Sapiens Duodenum Adenocarcinoma (HUTU-80), Human Embryonic Kidney (HEK-293) and Madin-Darby Canine Kidney (MDCK-II) cells chosen for their well-known physiology and wide use in toxicology experiments.[20–22] Among these cell lines, MDCK-II are known to form tight-junctions leading to barrier tissue formation, while HUTU-80 and HEK-293 cell cultures do not display barrier properties, despite the formation of densely packed adherent layers.[23] Cell presence was evaluated by acquiring transfer characteristic curves at periodic intervals of approximately 2-4 hours, for a total timeframe of 3 days. For the entire duration of the experiments, both the MNT and CTL devices were maintained in the cell incubator immersed in cell culture medium (Dulbecco's modified Eagle's medium, DMEM), along with control substrates for assessing cell proliferation dynamics. To this end, a well-established viability assay based on the MTT reagent (3-(4,5-dimethylthiazol-2-yl)-2,5-diphenyltetrazolium bromide) was used.[24,25]

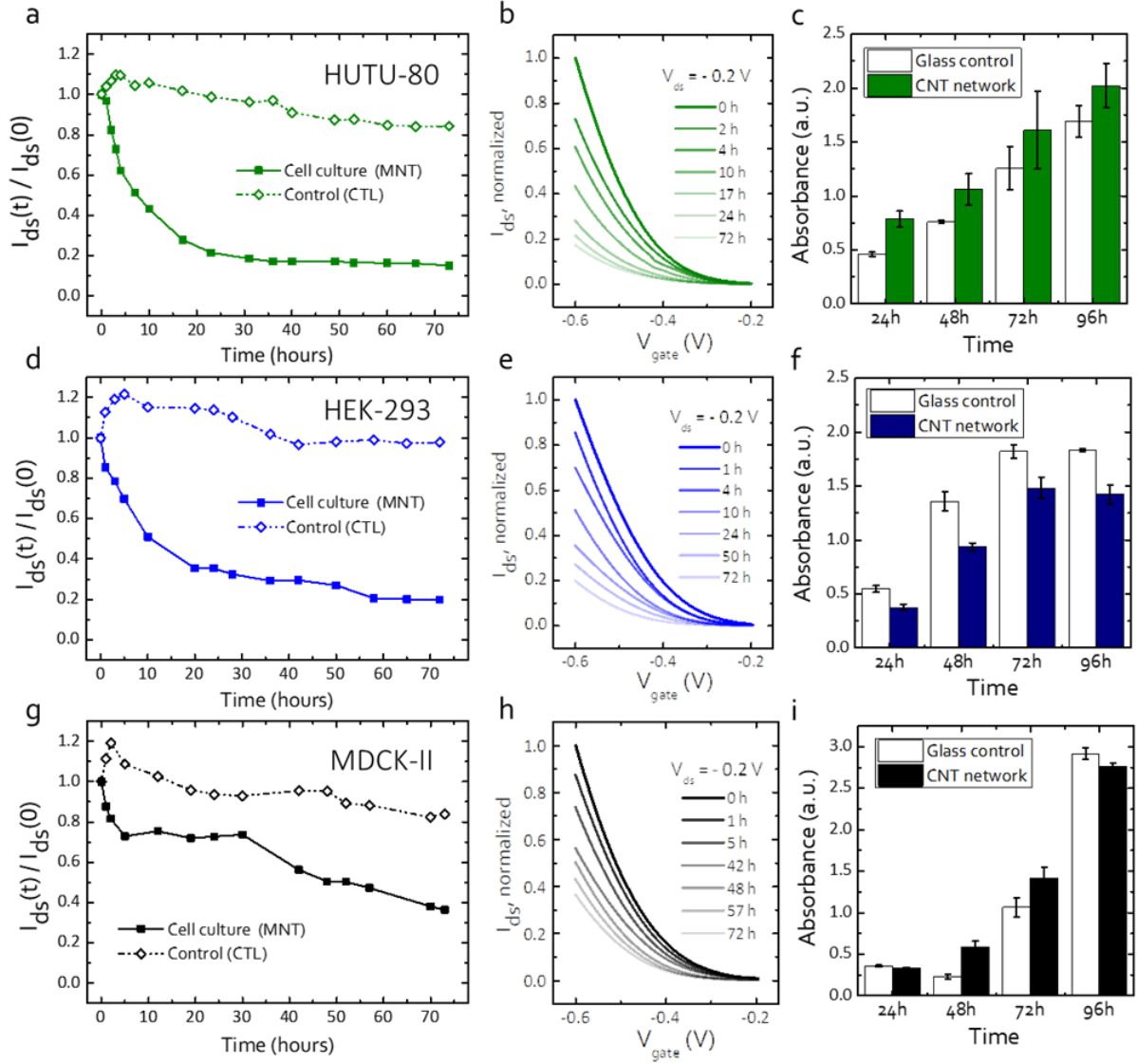

**Figure 2.** Cell adhesion and proliferation monitoring. Summary of the cell viability monitoring performed on three different cell lines: HUTU-80 (green), HEK-293 (blue) and MDCK-II (black). (**a,d,g**) Relative variations in $I_{ds}$ for the three cell models, normalized to the initial measurement ($t = 0$ s): in all experiments the CTL devices (void diamonds) demonstrate a good stability throughout the measurements, while MNT currents (full squares) experience a decrease owing to the establishment of an added capacitive contribution in series to $C_{EDL}$. (**b,e,h**) Transfer characteristic curves of the MNT devices in linear regime, acquired at different times during the proliferation experiment, showing that $I_{ds}$ drops as a consequence of lower effective gate

capacitance rather than $V_{th}$ shifts (see Supporting Information Figure S6 for CTL). **(c,f,i)** MTT cell viability assays performed alongside the electrical measurements on CNT networks (full) and glass controls (void), demonstrating cell viability of the CNT layer and providing indications on proliferation dynamics of the individual cellular models (average value over 3 samples for each point, error bars indicate the standard deviation).

For all the considered cellular models, the time evolution of the MNT current, owing to the decrease in the slope of the devices transfer characteristic curves, is in agreement with the reduction in effective capacitive coupling at the semiconductor/electrolyte interface (Figure 2). This can indeed be ascribed to the additional contribution of the cell layers in series to the EDL capacitance. In particular $C_{cl}$ is likely the compound contribution of the cleft region and of the whole cell body.[3] The evolution of the $I_{ds}$ currents in time can also provide qualitative information on the proliferation dynamics. In fact, the behaviour of the MNT currents for HEK-293 and MDCK-II cell lines are in good agreement with the results of the MTT tests. In the case of HEK-293 proliferation, the onset of a plateau, ascribable to cells reaching confluence on the substrates, occurs 60 h after seeding, while a slower dynamic characterizes the first two days of proliferation of the MDCK-II. In the case of the HUTU-80 cell lines such agreement is weaker, with the current in the MNT device that appears to stabilize much earlier than what expected from the MTT assay. Nonetheless, the trend is respected also in this case, and different realizations of the MNT device show better agreement with the MTT assay (see Figure S10), suggesting a possible effect of non-homogeneous spatial distribution of cells on the substrate during the seeding/adhesion phase, thus highlighting the local nature of the information that our sensor provides on cellular adhesion and proliferation processes, in contrast to the macroscopic information given by MTT. Such an effect implies that more than one realization is necessary to extract the real cell proliferation analysis, an obvious conclusion in the light of unavoidable experimental noise and variability. Overall, the strength of the proposed approach is demonstrated by the consistency of the proliferation trends recorded by MNT devices across all of the three cell lines, as well as by the general reproducibility of the results with reiterations of the same experiments (see Figure S10), with respect to CTL devices displaying only minor deviations over the 3 days period, barely exceeding 20% of the initial value in the HEK-293 case.

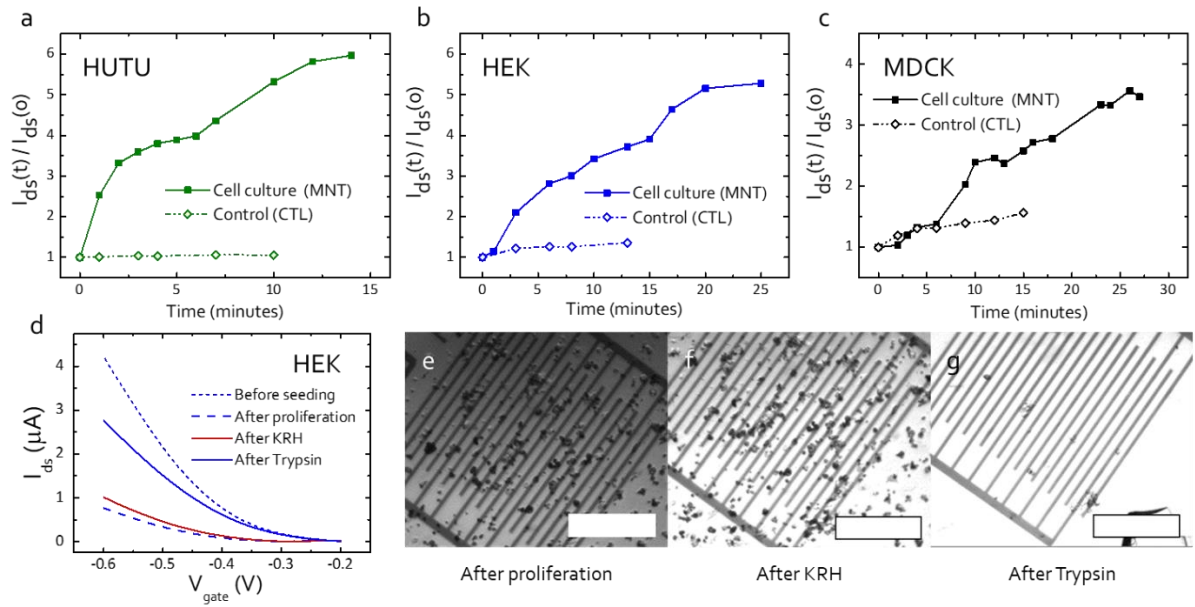

**Figure 3.** Cell detachment monitoring. (**a-c**) Relative variations in $I_{ds}$ for the three cell models, normalized to the initial measurement before Trypsin administration ($t = 0$ s): in all experiments the MNT devices (full squares) experience an increase in current demonstrating reversibility of the sensing mechanism, while CTL devices (void diamonds) maintain consistent current values with only minor variations compared to MNTs. (**d**) Transfer characteristic curves of the MNT device (HEK-293) in linear regime, acquired at different moments: before cell seeding (short-dash), after proliferation (dash), after electrolyte substitution (solid red), after Trypsin

administration (solid blue). (**e-g**) Microscope images showing cell presence on the device active area after proliferation (**e**) and after electrolyte substitution (**f**), while cells are absent after Trypsin treatment (**g**) (scale bars corresponding to 500 μm).

To validate the trends obtained during cell adherence and proliferation, the cell layers in the MNT devices were detached from the semiconductor/electrolyte interface via Trypsin treatment, with the CTL being exposed to the same treatment. Figure 3a-c displays the evolution of the $I_{ds}$ current levels for both MNT and CTL devices, relative to the initial values prior to Trypsin administration. As expected from the proposed working mechanism, MNT currents increase as cell progressively detach from the semiconductor/electrolyte interface, demonstrating the reversibility of the sensing mechanism, while the CTL devices are only subject to minor current increase possibly due to electrolyte-specific effects. The transfer characteristic curves (Figure 3D) further highlight how the variation in $I_{ds}$ is ascribable to changes in the capacitive coupling, as the slope of the curves decreases after proliferation, and it is essentially restored upon cell detachment, with differences possibly due either to a minor loss in perfomance after continued operation in aqueous environment (exceeding 72 hours) or to electrolyte-specific effects. To ensure the reduced impact of the latter on device behaviour, prior to Trypsin treatment the device was exposed to a different electrolyte with similar ionic content, i.e. Krebs-Ringer-HEPES buffer (KRH). As visible in Figure 3D, the change in gating medium does not significantly alter the transfer curves of the cell-covered device (dash-blue line vs. solid-red).

Furthermore, it is interesting to note that the current trends reported so far for both cells adhesion and detachment have been evaluated starting from the transfer curves acquired periodically throughout the measurements. Nonetheless, the sensing mechanism of these devices allows also for an alternative and simpler data acquisition approach, consisting in the continuous monitoring of current levels at a selected bias point for the entire duration of the experiment (see Supporting Information Figure S11).

**Conclusions**

In summary, EGFETs based on solution-processed SWCNT networks allow to electrically monitor cell adhesion/detachment processes *in vitro* by exploiting electrostatic interactions occurring at the cell/semiconductor interface. Such interactions decrease the strong capacitive coupling that is established at the semiconductor/electrolyte interface upon gate biasing. The key advantage of this approach with respect to other emerging technologies is the possibility of monitoring cell adhesion by simply evaluating the modifications of the device quasi-static current-voltage characteristics over the period of cells proliferation, without any need for electrochemical spectroscopy, or other frequency-dependent techniques, which in turn require complex signal processing and data analysis. This fundamental difference, together with the remarkable operating stability of the devices, their compatibility with immuno-fluorescence staining microscopy, and the ease of processability of the active materials, which are compatible with large-scale depositions through printing techniques, opens the path to a novel set of tools for bioviability monitoring with enhanced potential for automation, parallelisation and portability. Future efforts will focus on the quantitative assessment of cell coverage dynamics using this technique, along with the possibility of performing ECIS with the SWCNT EGFETs, thus expanding the analytical tools and the pervasivity of this cell-sensing platform.

**Materials and Methods**

*SWCNTs dispersion preparation:* Poly(3-dodecylthiophene-2,5-diyl) was synthesized via GRIM method.[26] The molecular weight ($M_n$ = 19.200 g mol$^{-1}$, $M_w$ = 22.300 g mol$^{-1}$, polydispersity index PDI: 1.16) was determined by gel permeation chromatography (GPC). HiPCO CNTs were purchased from Unidym, Inc. and were used as received. For the selection of SWNTs a previously reported procedure was modified.[14,27] The polymer was solubilized in toluene using a high power ultrasonicator (Misonix 3000) with cup horn bath (output power 69 W). Subsequently, CNTs were added to form the HiPCO:polymer dispersions with weight ratio 1:2. The solution was then sonicated for 2 h at 78 W and 12 °C. After ultrasonication, the dispersion was centrifuged at 30 000 rpm (109 000 g) for 1 h in an ultracentrifuge (Beckman Coulter Optima XE-90; rotor: SW55Ti) to remove all the remaining bundles and heavy-weight impurities. After the centrifugation, the highest density components precipitate at the bottom of the centrifugation tube, while the low-density components, including individualized semiconducting SWCNTs wrapped by the polymer and free polymer chains, stay in the upper part as supernatant. An extra step of ultracentrifugation was implemented to decrease the amount of free polymer in solution (enrichment).[28] To this purpose, the supernatant obtained after the first ultracentrifugation was centrifuged for 5 h, 55 000 rpm (367 000 g), the individualized SWCNTs were now precipitated to form a pellet and the free polymer was kept in the supernatant. Finally, the pellet was re-dispersed by mild sonication in o-xylene unless otherwise indicated.

*Samples and devices fabrication:* Low alkali 1737F Corning glasses were used as substrates for films and devices realized in this work. A standard cleaning in ultrasonic bath of Milli-Q water, acetone and isopropyl alcohol respectively and a following exposition to $O_2$-plasma at 100 W were employed. Bottom electrodes were patterned by a lift-off photolithographic process and deposited by evaporation of a 1.5 nm thick Cr adhesion layer and 15 nm thick Au film. Channel width *W* and length *L* are, respectively, 20 mm and 40 μm. Patterned substrates were cleaned by ultrasonic bath in isopropyl alcohol for 2-3 min and exposed to $O_2$ plasma at 100 W for 10 min prior to the surface treatment with hexamethyldisilazane (HMDS) vapours in $N_2$ atmosphere. After the deposition of the self-assembled monolayer SWCNT dispersions in o-xylene were spun on substrates at 1000 rpm for 90 s.

*Film characterization:* The surface topography of the SWCNT film films was measured with an Agilent 5500 Atomic Force Microscope operated in the Acoustic Mode.

*Cell cultures preparation:* Homo Sapiens Duodenum Adenocarcinoma (HUTU-80) and Human Embryonic Kidney (HEK-293) cells were acquired from American Type Culture Collection; Madin-Darby Canine Kidney (MDCK-II) cells were purchased from Sigma-Aldrich. Cells were cultured in cell culture flasks containing Dulbecco's modified Eagle's medium (DMEM) with 10% fetal bovine serum, 100 µg mL$^{-1}$ streptomycin, 100 U mL$^{-1}$ penicillin, and 100 U mL$^{-1}$ l-glutamine. Culture flasks were maintained in humidified atmosphere at 37 °C with 5% $CO_2$. When at confluence, the cells were enzymatically dispersed using Trypsin-EDTA and then plated on the SWCNT EGFETs and on glass control substrates for MTT viability assays.

*Cell-proliferation monitoring set-up and measurements:* Wells for cell proliferation were fabricated by 3D printing (Invicta 915, XFAB) and attached to EGFET substrate by means of paper clips, with butyl rubber o-rings ensuring tightness; both wells and o-rings were coated with approximately 3 µm of parylene to avoid interaction with cell culture medium. Assembled samples were sterilized by temperature treatment at 120 °C for 2 hours. Samples were mounted in a common electrical joint box hosting the required connections, which were sterilized by an ethanol bath (70%).

The proliferation phase was monitored while keeping the set-up in a cell incubator (Series II Water Jacket, Thermo Scientific) at controlled temperature and atmosphere (37 °C, 5% $CO_2$, humidified), while the detachment was performed in air.

The experimental protocol adopted during the proliferation phase is the following: (i) after complete sterilization of the components and set-up preparation, two EGFETs undergo a 30 minutes bias stress with DMEM as gating medium to condition the devices, after which simultaneous transfer curves are acquired on both devices. The EGFET presenting the larger $I_{ds}$ modulation is chosen as MNT, while the other acts as control. (ii) A fraction of DMEM is removed from the MNT well, and cells are subsequently seeded (approximately $10^4$ cells ml$^{-1}$, unless otherwise specified). An initial transfer characteristic curve is acquired, and the monitoring process is subsequently started. (iii) The set-up is inserted and maintained in a cell incubator for the entire duration of the measurements, accounting to approximately 72 hours, $I_{MNT}$ and $I_{CTL}$ levels being monitored via transfer characteristic curves acquisition. The current values reported in the Figures refer to a $V_{gate}$ = -0.6 V, $V_{ds}$ = -0.2 V voltage bias. In continuous monitoring mode, devices were measured at a constant bias of -0.4 V and -0.2 V for $V_{gate}$ and $V_{ds}$ respectively, sampling $I_{ds}$ currents every 10 s in the proliferation phase, and every 1 s in the detachment phase.

To further verify the validity of the approach, the monitoring process is performed also during cell detachment from the SWCNT network, induced by substituting in both MNT and CTL samples the cell culture medium with trypsin 2.5%. While changing the gating medium, the gate electrode is again exposed to air because of the set-up geometry and configuration. A semiconductor device analyzer (Agilent B1500A) was used for the electrical measurements.

*Optical cell viability assay and immunofluorescence staining:* The proliferation was evaluated after 1, 2, 3, and 4 d in vitro. For each time point the medium was removed and replaced with RPMI without phenol red containing 0.5 mg mL$^{-1}$ of MTT reagent (3-(4,5-dimethylthiazol-2-yl)-2,5-diphenyltetrazolium bromide, Sigma-Aldrich). Cells were reincubated at 37 °C for 3 h. Formazan salt produced by cells through reduction of MTT was then solubilized with 200 mL of ethanol and the absorbance was read at 560 and 690 nm (using a microplate reader TECAN Spark10M). The proliferation cell rate was calculated as the difference in absorbed intensity at 560 and 690 nm.

Cells grown on glass coverslips coated with SWCNT networks were washed twice with PBS and fixed for 15 min at RT in 4 % paraformaldehyde and 4 % sucrose in 0.12 M sodium phosphate buffer, pH 7.4. Fixed cells were pre-incubated for 20 min in gelatin dilution buffer (GDB: 0.02 M sodium phosphate buffer, pH 7.4, 0.45 M NaCl, 0.2% (w/v) gelatin) containing 0.3% (v/v) Triton X-100, and subsequently incubated with Phalloidin Alexa Fluor 488 conjugated in GDB for 1 h at RT and finally washed with PBS. The images were acquired with an inverted fluorescence microscope (Nikon Eclipse Ti-U equipped with LED sources, Lumencor Spectra X).

AUTHOR INFORMATION

**Corresponding Author**

* E-mail: Mario.Caironi@iit.it

**Present Addresses**

† Department of Biomedical Engineering, Tufts University, Medford, MA 02155, USA.

**Author Contributions**

‡ F. Scuratti and G.E. Bonacchini contributed equally. F. Scuratti and G.E. Bonacchini contributed to experiment design, measurements, data analysis, and manuscript preparation. C. Bossio performed the optical cell viability assays and contributed to the EGFET-based cell proliferation and detachment measurements. J.M. Salazar-Rios, W. Talsma, M.A. Loi prepared the SWCNT dispersions. M.R. Antognazza and M. Caironi supervised the work and contributed to planning experiments, data analysis, and manuscript preparation.

**Funding Sources**

M. C. acknowledges support by the European Research Council (ERC) under the European Union's Horizon 2020 research and innovation program 'HEROIC', grant agreement 638059. M. A. acknowledges support by the European Research Council (ERC) under the European Union's Horizon 2020 research and innovation program 'LINCE', grant agreement 803621.

**Notes**

The authors declare no competing financial interest.

ACKNOWLEDGMENT

The authors wish to acknowledge Luca Frezza for his contribution to the design and fabrication of the 3D printed wells, Elena Stucchi and Paolo Colpani for the parylene coating process, Aprizal Akbar Sengrian for the polymer-wrapped SWCNT sample preparation, as well as Ullrich Scherf and Sybille Allard. This work has been partially carried out at Polifab, the micro- and nano-technology center of the Politecnico di Milano.

https://doi.org/10.1016/J.TOX.2010.03.017.

(22) Kroll, A.; Dierker, C.; Rommel, C.; Hahn, D.; Wohlleben, W.; Schulze-Isfort, C.; Göbbert, C.; Voetz, M.; Hardinghaus, F.; Schnekenburger, J. Cytotoxicity Screening of 23 Engineered Nanomaterials Using a Test Matrix of Ten Cell Lines and Three Different Assays. *Part. Fibre Toxicol.* **2011**, *8* (1), 9. https://doi.org/10.1186/1743-8977-8-9.

(23) Ramuz, M.; Hama, A.; Rivnay, J.; Leleux, P.; Owens, R. M. Monitoring of Cell Layer Coverage and Differentiation with the Organic Electrochemical Transistor. *J. Mater. Chem. B* **2015**, *3* (29), 5971–5977. https://doi.org/10.1039/C5TB00922G.

(24) Gerlier, D.; Thomasset, N. Use of MTT Colorimetric Assay to Measure Cell Activation. *J. Immunol. Methods* **1986**, *94* (1–2), 57–63. https://doi.org/10.1016/0022-1759(86)90215-2.

(25) Mosmann, T. Rapid Colorimetric Assay for Cellular Growth and Survival: Application to Proliferation and Cytotoxicity Assays. *J. Immunol. Methods* **1983**, *65* (1–2), 55–63. https://doi.org/10.1016/0022-1759(83)90303-4.

(26) Loewe, R. S.; Khersonsky, S. M.; McCullough, R. D. A Simple Method to Prepare Head-to-Tail Coupled, Regioregular Poly(3-Alkylthiophenes) Using Grignard Metathesis. *Adv. Mater.* **1999**, *11* (3), 250–253. https://doi.org/10.1002/(SICI)1521-4095(199903)11:3<250::AID-ADMA250>3.0.CO;2-J.

(27) Gomulya, W.; Costanzo, G. D.; de Carvalho, E. J. F.; Bisri, S. Z.; Derenskyi, V.; Fritsch, M.; Fröhlich, N.; Allard, S.; Gordiichuk, P.; Herrmann, A.; et al. Semiconducting Single-Walled Carbon Nanotubes on Demand by Polymer Wrapping. *Adv. Mater.* **2013**, *25* (21), 2948–2956. https://doi.org/10.1002/adma.201300267.

(28) Bisri, S. Z.; Gao, J.; Derenskyi, V.; Gomulya, W.; Iezhokin, I.; Gordiichuk, P.; Herrmann, A.; Loi, M. A. High Performance Ambipolar Field-Effect Transistor of Random Network
22

Carbon Nanotubes. *Adv. Mater.* **2012**, *24* (46), 6147–6152. https://doi.org/10.1002/adma.201202699.

# Supporting Information

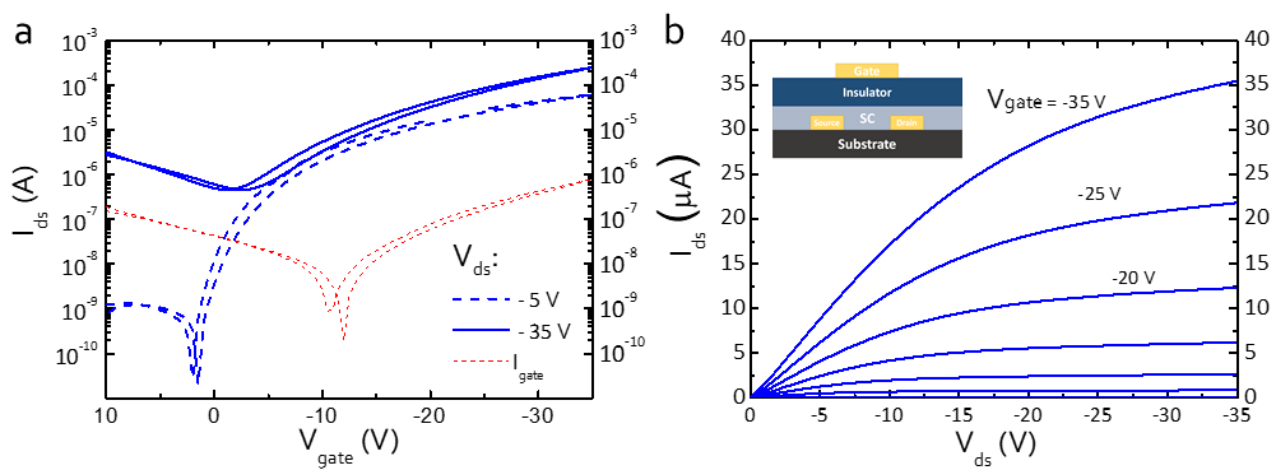

**Fig. S1. Solid-state FET structure and operation.** Transfer (**a**) and output (**b**) characteristic curves of the solid-state FET, with schematic diagram of the top-gate bottom-contact device (**b**, insert). In the linear regime the ON/OFF ratio reaches $10^5$ at a relatively low driving voltage of 30V. In the saturation region, the onset of device ambipolarity limits the ON/OFF ratio to approximately $10^2$. In the linear and saturation regimes, charge carrier mobilities are in the range of 0.49 and 0.73 cm$^2$V$^{-1}$s$^{-1}$ which, although not exhibiting the full potential of the material[1], are still notable values since devices have not been further optimized. Threshold voltages are respectively -24 V and -14 V for the linear and saturation regimes, and device operation is not significantly affected by hysteresis. Finally, from the shape of the output curve at low $V_{ds}$, a mild effect of contact resistance on device behaviour is visible.

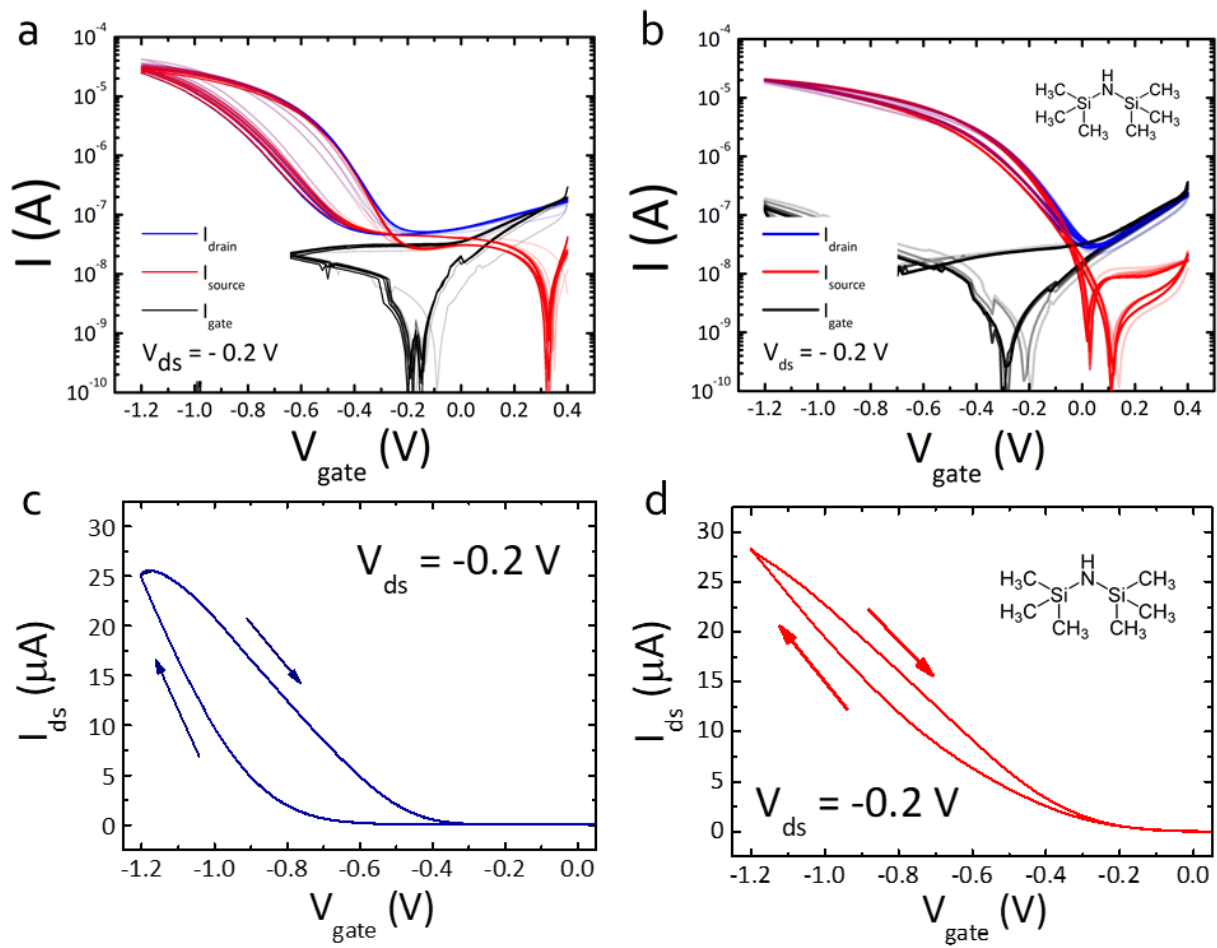

**Fig. S2. Impact of the substrate surface treatment.** Prior to CNT network deposition, hexamethyldisilazane (HMDS) treatment was performed to improve EGFET performances and operating stability in aqueous environment. (**a,b**) Series of ten transfer characteristic curves (semi-log scale) acquired on an untreated (**a**) and a HMDS-treated (**b**) devices operating in ultrapure water, the former showing a large hysteresis increasing with the number of acquisitions, while the latter presents limited hysteresis, negligible variations in performances, and reduced threshold voltage (inset, molecular structure of HMDS). (**c,d**) Linear scale plots of the transfer characteristics for devices without and with HMDS treatment, the latter (**d**) presenting a more ideal behaviour compared to the untreated device (**c**).

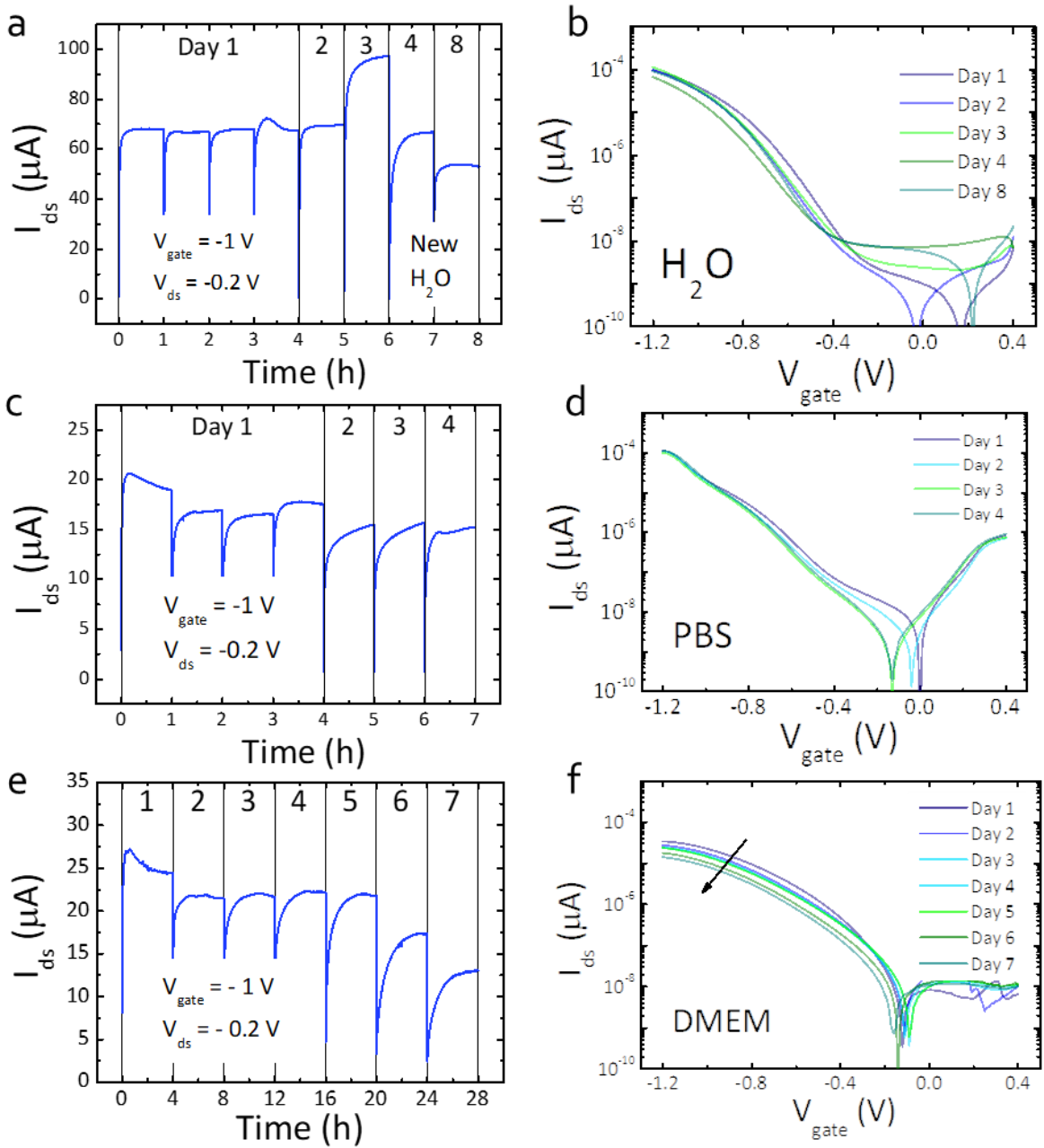

**Fig. S3. Operating stability of EGFETs in aqueous environments.** Bias stress tests (**a,c,e**) and transfer characteristic curves (**b,d,f**) of SWCNT (spun from ortho-dichlorobenzene) based EGFETs operating in three different gating media. Bias-stress tests are performed by monitoring the temporal evolution of the $I_{ds}$ current upon constant biasing, at fixed $V_{ds}$ and $V_{gate}$. Transfer curves are usually measured at the beginning and after the bias-stress, to decouple threshold voltage variations from other possible effects. The experimental results hereby reported have been obtained by running measurements for several hours per day, for few days up to one week. Three separate devices have been exploited for measuring operating behaviour in ultrapure water, phosphate-buffered saline (PBS) and Dulbecco's modified Eagle's medium (DMEM) respectively; for the whole duration of the experiments, i.e. from four to eight days, the channel area has been constantly maintained in liquid environment by adding gating medium to compensate for its evaporation. (**a,b**) The bias-stress test in pure water has been performed by applying constant bias for intervals of 60 minutes, for a total of 4 h during the first day, 1 h during the second, third, fourth and eighth day each. The $I_{ds}$ current value of ≈68 µA is constant during the experiment, with the exception of day 3 data, in which a marked increase exceeding 42% is visible. Such rather large and unexpected variation is due to water contamination during sample storage, as demonstrated by the recovery of the initial current level after the gating medium was replaced with new water - ensuring that the channel area was not exposed to air during the process. A decrease in current (-22%) is also visible in the last section of the curve, representing device operation after 8 days being kept continuously immersed. (**c,d**) Identical measurements have been carried out by adopting PBS as gating electrolyte. Similarly to the previous case, after an initial adjustment the device regularly provides a current level of approximately 15 µA. It can be noted that after the first day the transient current dynamics slightly change, and apparently it does not reach a stationary plateau within the first 60 minutes. This could be ascribed to a variation in the concentration of solutes in the electrolyte, as small amounts of fresh solution have been added daily to compensate for evaporation of the gating medium. To be noted that the transfer characteristics of the device used in these measurements are not ideal, the curve showing an unexpected change in slope around 1 V. (**e,f**) With respect to water and PBS, the bias-stress test carried in DMEM has been performed for more prolonged time intervals, i.e. 4 hours, with measurements carried out every day for an entire week, in line with the expected timeframes for proliferation of most largely used model cell lines. After an initial adjustment, $I_{ds}$ settles at 22 µA, with significant variations occurring only on the sixth and seventh day (23 and 45%, respectively). These variations are ascribable to a shift in the threshold voltage to more negative values, as indicated by the transfer characteristics of the device, and it is in general compatible with effects commonly induced by bias stressing. To be noted that in current literature there are very few examples of such stability stress tests performed on EGFETs, the most recent and significant being provided in Ref. 15,30. To the best of our knowledge, no EGFET has demonstrated comparable operating stability for such long periods of time, maintained in diverse aqueous environments.

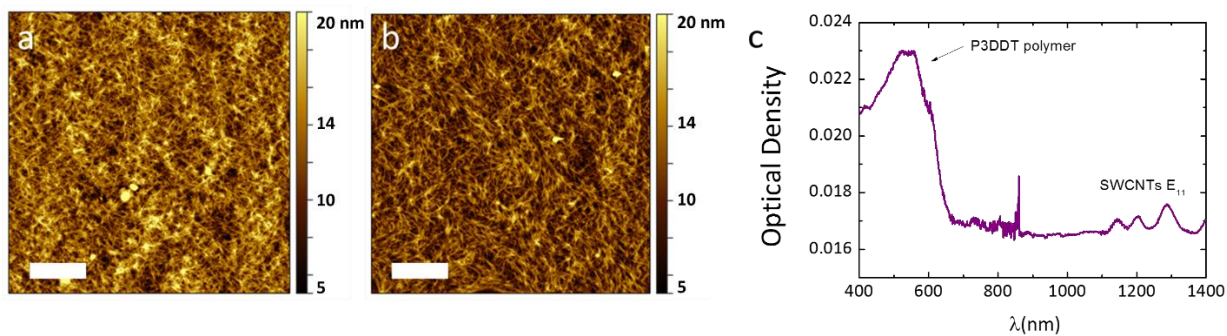

**Fig. S4. Thin film characterization of CNT networks.** (**a,b**) AFM images of CNTs spin-coated on a glass substrate treated with HMDS, as deposited (**a**) and after 3 days of immersion in ultrapure water, dried before measurement (**b**). In both cases, uniform and well inter-connected networks of CNT are visible, with a thickness that in general does not exceed 10 nm, and a surface roughness of 2.55 and 2.47 nm rms respectively. The good coverage in the range of 90-95% highlights the optimal loading of CNT dispersed in the solvent. Scale bars: 1 µm. (**Cc**) Absorption spectrum of spin-casted HiPCO:P3DDT SWCNTs on glass, showing the low optical density of the films.

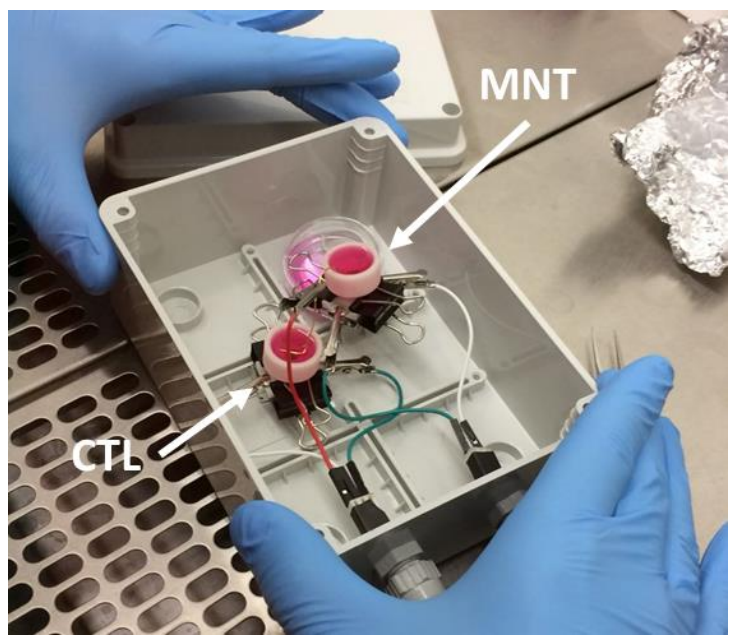

**Fig. S5. Measurement set-up.** Picture of the set-up comprising the monitoring (MNT) and control (CTL) devices, the wells for containing the cell culture medium, and the electrical connections.

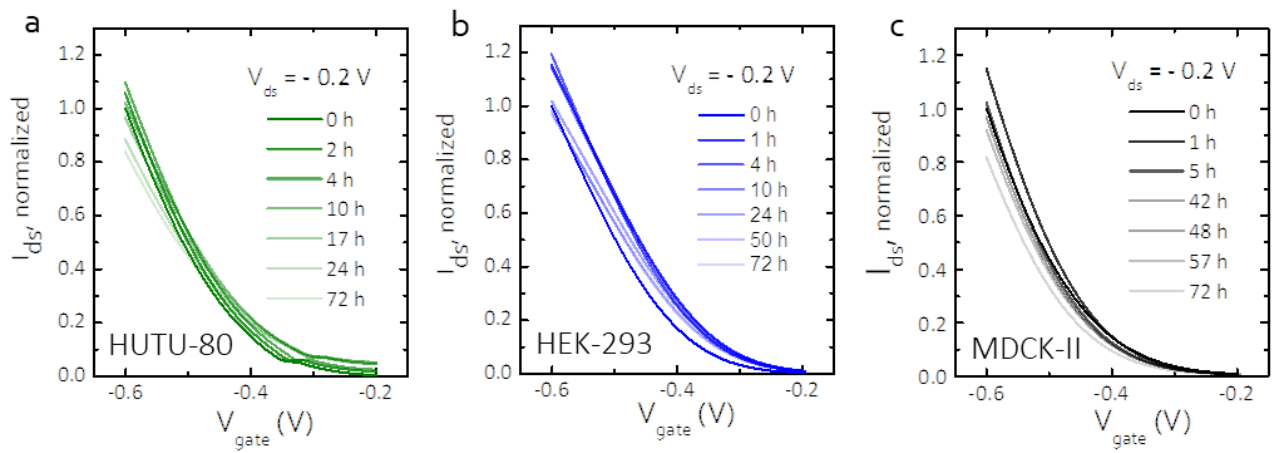

**Fig. S6. Cell adhesion and proliferation monitoring, control experiment.** (**a,b,c**) Transfer characteristic curves of the CTL devices in linear regime, acquired at different moments during the proliferation experiment, showing minor oscillations of $I_{ds}$.

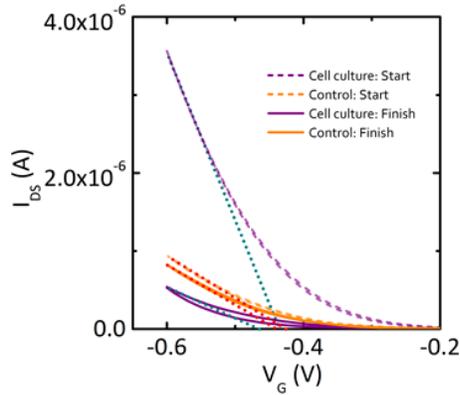

**Fig. S7. Transfer curve analysis – HUTU-80.** Transfer characteristic curves of MNT (purple) and CTL (orange) devices before (dash) and after (solid) cell proliferation. Table reports slope and $V_{th}$ values for each curve.

|  | Slope | $V_{th}$ |
|---|---|---|
| Cell: start | $-2.23522 \times 10^{-5}$ | $-0.4357$ |
| Cell: finish | $-3.9913 \times 10^{-6}$ | $-0.4634$ |
| Control: start | $-5.61141 \times 10^{-6}$ | $-0.4262$ |
| Control: finish | $-4.99906 \times 10^{-6}$ | $-0.4415$ |

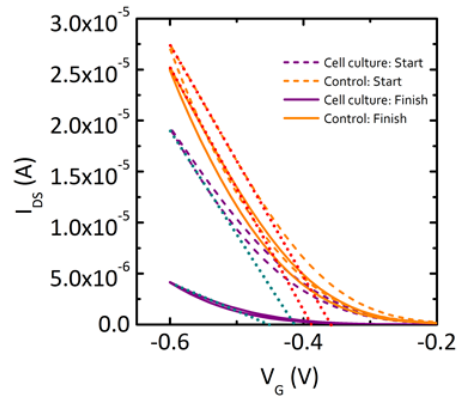

**Fig. S8. Transfer curve analysis – HEK-293.** Transfer characteristic curves of MNT (purple) and CTL (orange) devices before (dash) and after (solid) cell proliferation. Table reports slope and $V_{th}$ values for each curve.

|  | Slope | $V_{th}$ |
|---|---|---|
| Cell: start | $-1.08826 \times 10^{-4}$ | $-0.4166$ |
| Cell: finish | $-3.00907 \times 10^{-5}$ | $-0.4629$ |
| Control: start | $-1.13881 \times 10^{-4}$ | $-0.3584$ |
| Control: finish | $-1.21211 \times 10^{-4}$ | $-0.3883$ |

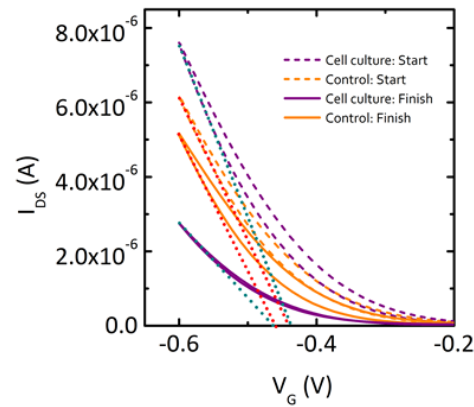

**Fig. S9. Transfer curve analysis – MDCK-II.** Transfer characteristic curves of MNT (purple) and CTL (orange) devices before (dash) and after (solid) cell proliferation. Table reports slope and $V_{th}$ values for each curve.

|  | Slope | $V_t$ |
|---|---|---|
| Cell: start | $-4.71899 \times 10^{-5}$ | $-0.4368$ |
| Cell: finish | $-1.8267 \times 10^{-5}$ | $-0.4629$ |
| Control: start | $-3.96428 \times 10^{-5}$ | $-0.4398$ |
| Control: finish | $-3.56024 \times 10^{-5}$ | $-0.4595$ |

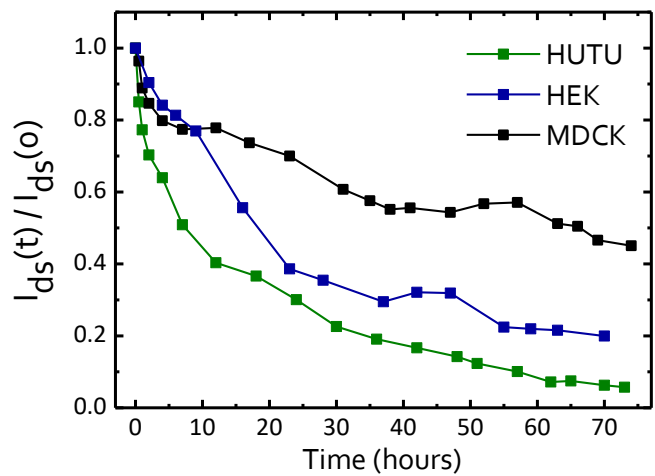

**Fig. S10. Cell adhesion trends.** Relative variations in $I_{ds}$ for the three cell models, normalized to the initial current ($t = 0$ s) performed in other sets of experiments, in order to test the reliability of the measurements. As for the trends reported in the main text (and in Fig. 2), the responses of the MNT devices demonstrate proliferation dynamics consistent with the information provided by the MTT assays. Note that in this set of experiments, the agreement between MTT assay and MNT device holds also for the HUTU-80 cell line (green).

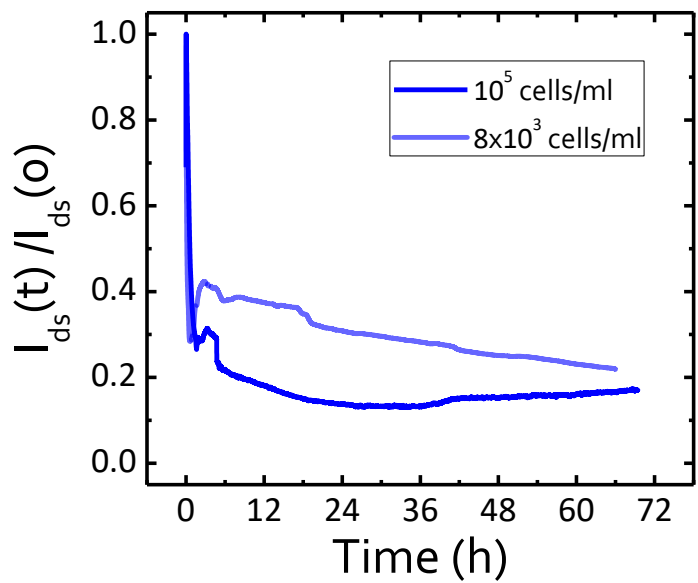

**Fig. S11. Continuous monitoring of cell adhesion and proliferation – HEK-293.** $I_{ds}$ trends of two MNT devices seeded with different initial concentrations of cells. The gate voltage is kept fixed at $V_{gate}$ = -0.4 V, with $V_{ds}$ = -0.2 V.

**Materials and Methods**

*CNT FET device fabrication and electrical characterization*
Reference, solid-state CNT FET devices were realized in bottom-contact top-gate configuration. The SWCNTs dispersion was spun from ortho-dichlorobenzene at 1000 rpm for 90 s and subsequently annealed at 150°C for 1 h. A dielectric layer with thickness between 500–600 nm was obtained by spin-casting Polystyrene (Sigma-Aldrich) with $M_w$ = 290 kg mol$^{-1}$ from n-butyl acetate (60 g L$^{-1}$). After the dielectric deposition, the devices were annealed under nitrogen, on a hot-plate, at 80 °C for 30 min. The transistors were then completed by evaporating a 40nm thick aluminium gate.
The electrical characteristics of solid-state FETs were measured in a nitrogen glovebox on a Wentworth Laboratories probe station with a semiconductor device analyser (Agilent B1500A). The linear and saturation mobility values were calculated using the gradual-channel approximation, according to the parallel plate model. The mobility data reported do not take into account contact effects.

*EGFET electrical characterisation*
The electrical characteristics and bias-stress monitoring of EGFETs were assessed in air with on a Cascade Microtech probe station with a semiconductor device analyser (Agilent B1500A), with an Au filament acting as gate electrode.

*Film characterization*

The surface topography of the films was measured with an Agilent 5500 Atomic Force Microscope operated in the Acoustic Mode.

Optical measurements were carried out on a Perkin Elmer λ1050 spectrophotometer, using a tungsten lamp as the excitation source.